\documentclass[aps,showpacs,amsmath,amssymb,pra,floatfix,superscriptaddress,a4,twocolumn]{revtex4}

\usepackage[dvips]{graphicx}
\usepackage{amsmath}
\usepackage{color}

\def\CR{\nonumber\\[0.15cm]}
\newcommand{\bra}[1]{\langle\,{#1}\, |}
\newcommand{\ket}[1]{|\,{#1}\,\rangle}
\newcommand{\braket}[2]{\mbox{$\langle\,{#1}\, | \,{#2}\,\rangle$}}

\newcommand{\rref}[1]{Ref.~\cite{#1}}
\newcommand{\fref}[1]{Fig.~\ref{#1}}
\newcommand{\frefp}[2]{Fig.~\ref{#1}~(#2)}
\newcommand{\bref}[1]{(\ref{#1})}
\newcommand{\eref}[1]{Eq.~(\ref{#1})}
\newcommand{\sref}[1]{section \ref{#1}}

\def\CR{\nonumber\\[0.15cm]}
\newcommand{\sub}[2]{{#1}_{\mbox{\!\! \scriptsize #2}}}
\newcommand{\bv}[1]{\mathbf{ #1 }}

\definecolor{orange}{rgb}{0.8,0.4,0.0}

\newcommand{\gstate}{g}

\newcommand{\sstate}{s}
\newcommand{\pstate}{p}
\newcommand{\Ntot}{2N}
\newcommand{\Pos}{r}

\begin{document}

\title{Source of entangled atom pairs on demand, using the Rydberg blockade}

\author{S. W{\"u}ster}
\affiliation{Max-Planck-Institute for the Physics of Complex Systems, 01187 Dresden, Germany}

\author{S. M{\"o}bius}
\affiliation{Max-Planck-Institute for the Physics of Complex Systems, 01187 Dresden, Germany}

\author{M. Genkin}
\affiliation{Max-Planck-Institute for the Physics of Complex Systems, 01187 Dresden, Germany}

\author{A. Eisfeld}
\affiliation{Max-Planck-Institute for the Physics of Complex Systems, 01187 Dresden, Germany}

\author{J. M. Rost}
\affiliation{Max-Planck-Institute for the Physics of Complex Systems, 01187 Dresden, Germany}

\date{\today}

\begin{abstract}
Two ultracold atom clouds, each separately in a dipole-blockade regime, realize a source of entangled atom pairs that can be ejected on demand.
Entanglement generation and ejection is due to resonant dipole-dipole interactions, while van-der-Waals interactions are predominantly responsible for the blockade 
that ensures the ejection of a single atom per cloud. A source of entangled atoms using these effects can operate with a $10$kHz repetition rate producing ejected atoms with velocities of about $0.5$m/s. 
\end{abstract}

  

\maketitle

\section{Introduction}\label{introduction}

The interaction between alkali atoms excited to Rydberg states is about $10$ orders of magnitude stronger than in the ground state \cite{book:gallagher,Saff_10}. This can lead to a complete blockade of double excitation in a small volume \cite{urban:twoatomblock,gaetan:twoatomblock}, a drastic effect that can be exploited
to engineer quantum gates~\cite{jaksch:dipoleblockade,lukin:quantuminfo}, giant optical non-linearities~\cite{adams:giantkerr,sevilay:nonlocalnonlin} and ultimately single-photon sources~\cite{saffmann:singleatomsource,dudin:singlephot,peyronel:singlephot}. 

In a sufficiently tightly confined atomic gas, the blockade allows precisely one atom to carry a Rydberg excitation. This atom can then be selectively separated from the gas using external fields, hence the blockade has also been proposed to realize single atom sources~\cite{saffmann:singleatomsource}.  As we show here, the use of resonant dipole-dipole interactions for the ejection of atom pairs will generate entangled (Bell) pair states and also offers an elegant way to avoid perturbations of the parent atomic gas. 

Our setup allows a replication of the spin variant \cite{aharonov:bohm:EPR} of the Einstein-Podolsky-Rosen (EPR) paradox \cite{epr:EPR}, utilizing massive particles, for which case only few experimental realizations exist, see e.g.~\cite{rowe:ionEPR}. Other proposals to study EPR correlations with massive cold particles, such as \cite{karen:moldisEPR,lewisswan:EPRsensitivity} and experiments in similar directions,e.g~\cite{gross:homodyneBEC,perrin:entangle4wavemix} show that this is an active research field.

Beyond the possible EPR source, the present atom source may have applications for atom-laser outcoupling~\cite{robins:atomlaser:review,haine:entangledatomlaser}, nanolithography~\cite{nguyen:rydberg:nanolith} or controlled collisions in ultracold chemistry \cite{ospelkaus:ultracold_chemical}.

\begin{figure}[htb]
\centering
\includegraphics[width=0.99\columnwidth]{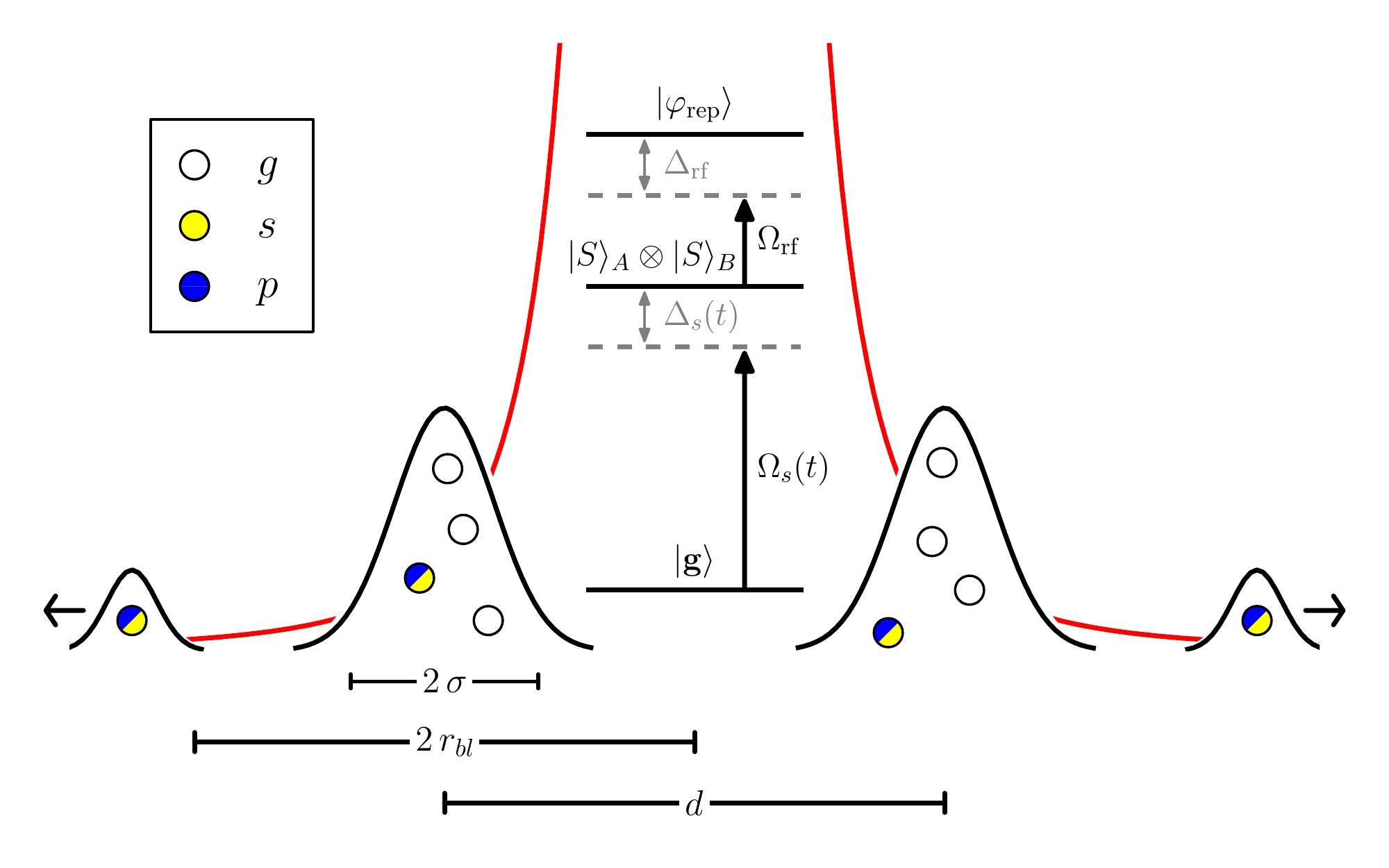}
\caption{(color online) Schematic of twin atom clouds. Shown are the Rydberg blockade radius $r_{bl}$, the cloud separation $d$ and width $\sigma$. The center shows the many-body level scheme and red lines symbolize the repulsive exciton potential. For simplicity $d$ here is slightly larger than $r_{bl}$, while for the scenario in \fref{simulation} it is slightly smaller, as explained in the text. The two-color circles symbolize atom pairs in superposition states $\ket{\Psi}=(\ket{sp} + \ket{ps})/\sqrt{2}$.
\label{system}}
\end{figure}
The essentials of our proposal are sketched in \fref{system}. The two tightly trapped atomic clouds have a separation $d$ slightly larger than the blockade radius $\sub{r}{bl}$ for a given principal quantum number $\nu$ of a Rydberg state. It is then possible to excite precisely one atom to a Rydberg state $\ket{\nu,l}$ in \emph{each} cloud, where $l$ is the orbital angular momentum. Initially we target the state $\ket{s}=\ket{\nu,l}$ with $l=0$.

Subsequently, a suitable microwave (r.f.) pulse on the $\ket{s}\rightarrow\ket{p}$ transition, where $p=\ket{\nu,l}$ with $l=1$, can excite the whole system to a collective state (exciton) with repulsive resonant dipole-dipole interactions between pairs of atoms from different clouds \cite{Mourachko:manybody,anderson:frozendipdip,mudrich:backandforth,park:dipdipionization,carrol:angulardipdip,Ditzhuijzen:Spatialdipdip}. At separation $d$, this interaction will be much stronger than van-der Waals interactions,
and accelerates the atoms away from one another \cite{cenap:motion,wuester:cradle,moebius:cradle}. In the repulsive exciton, the electronic state of the ejected atom pair is a Bell state \cite{wuester:cradle,moebius:cradle}, hence our setup directly implements the EPR scenario~\cite{aharonov:bohm:EPR}. In practice, it is beneficial to set up the clouds with $d\lesssim r_{bl}$. In that case the detuning of the Rydberg excitation lasers has to compensate for the van-der-Waals interaction at distance $d$, in an anti-blockade type setting \cite{cenap:antiblockade}.

This article is organized as follows: In \sref{model} we present our model and the methods needed for its solution. In particular we describe a modification of 
Tully's quantum-classical surface hopping method~\cite{tully:hopping2,tully:hopping}, that allows us to treat an explicitly time-dependent Hamiltonian and thus include the excitation and acceleration processes of the atoms in the same modelling scheme (\sref{numerics}).
In \sref{ejection} we use this method to simulate the ejection of an entangled atom pair from two traps, and determine key relations between system parameters and repetition rate.
Later, in \sref{entanglement}, we describe protocols to measure the violation of a Bell inequality~\cite{bell:theorem,chsh:inequality} in the present setting, which also allow 
an experimental verification of entanglement transport based on the same Rydberg Bell states as those occuring here~\cite{wuester:cradle,moebius:cradle}. 

\section{Model and methods}\label{model}

Consider an assembly of $N=N_A +N_B$ neutral atoms of mass $M$ located at positions $\Pos_n$, restricted to one dimension and confined in two separate harmonic wells. This could be realized by trapping atoms in two sites of an optical lattice, not necessarily adjacent, or a double well trap. $N_A$ atoms are localized in one of the wells, forming cloud $A$ and $N_B$ in the other well, forming cloud $B$. Near the centres of each well at $x=\pm d/2$, the potential is approximately harmonic: $V(\Pos_n)=M \omega^2 (\Pos_n\pm d/2)^2/2$, and the atoms are initially in the Gaussian ground state of the trap with width $\sigma=\sqrt{\hbar/M\omega}$. Under the blockade conditions described in the introduction, one expects that after Rydberg excitation only a single atom per trapping site undergoes significant motional dynamics, as we have shown previously \cite{moebius:bobbels}. If no real Rydberg states are accessed, but \emph{all} ground states are off-resonantly dressed with Rydberg states~\cite{moebius:bobbels}, the system can evolve into a spatial mesoscopically entangled state through dipole-dipole interactions \cite{moebius:cat,genkin:dressedbobbels}. The present paper however, does not consider the dressing scenario.

\subsection{Hamiltonian and state space}\label{hamil}

We consider three essential states in $^{87}$Rb atoms. A long lived ground state $\ket{g}$ and two Rydberg states $\ket{\nu,l}$, designated by $\ket{s}=\ket{\nu,0}$ and $\ket{p}=\ket{\nu,1}$. On these states we build the many-body basis $\ket{\bv{k}}\equiv \ket{k_1\dots k_{\Ntot}}  \equiv\ket{{k_{1}}} \otimes \dots \otimes\ket{k_{\Ntot}}$, where $k_{j}\in \{S\}\equiv\{{\gstate},{\sstate},{\pstate}\}$ describes the electronic state of atom $j$. 

The state $\ket{s}$ is coupled to the ground state with Rabi frequency $\sub{\Omega}{las}$ and detuning $\sub{\Delta}{las}$
via a two-photon laser transition, and to the state $\ket{p}$ with Rabi frequency $\sub{\Omega}{rf}$ and detuning $\sub{\Delta}{rf}$
via a one-photon microwave transition. The resulting coupling between many-body states is sketched in \fref{system} and will be explained shortly. 

We formulate the many-body Hamiltonian
\begin{subequations}
\begin{align}
\hat{H} &= \sub{\hat{H}}{0} + \sub{\hat{H}}{c} + \sub{\hat{H}}{int}, 
\label{Hamiltonian}
\\
 \sub{\hat{H}}{0} &=\sum_{n=1}^{\Ntot}\left[ -\frac{\hbar^2}{2 M} \nabla^2_{\Pos_n}  +  V(\Pos_n,t)\hat{\sigma}^{(n)}_{gg} \right],
\\
 \sub{\hat{H}}{c} &=\sum_{n}\bigg[\frac{\sub{\Omega}{las}(t)}{2}\hat{\sigma}^{(n)}_{gs}  + \frac{\sub{\Omega}{rf}(t)}{2}\hat{\sigma}^{(n)}_{sp}  + \mbox{h.c.} 
\CR
& -\sub{\Delta}{las}(t)\hat{\sigma}^{(n)}_{ss} -[\sub{\Delta}{rf}(t)+\sub{\Delta}{las}(t)]\hat{\sigma}^{(n)}_{pp}  \bigg],
\end{align}
\begin{align}
 \sub{\hat{H}}{int} &=\sum_{nl}\bigg[ D_{nl}\hat{\sigma}^{(n)}_{sp}\hat{\sigma}^{(l)}_{ps} + \sum_{a,b=s,p} W^{(ab)}_{nl}\hat{\sigma}^{(n)}_{aa}\hat{\sigma}^{(l)}_{bb}  \bigg],
\label{InteractionHamiltonian}
\end{align}
\end{subequations}
with $\hat{\sigma}^{(n)}_{kk'}=\ket{k_{n}}\bra{k_{n}'}$, where $k_n$, $k'_n$ $\in S$. 
The operator $\hat{\sigma}^{(n)}_{kk'}$ acts only on the Hilbert space of atom $n$ and is unity otherwise. In \eref{Hamiltonian}, $\sub{\hat{H}}{0}$ describes the free motion of the atoms with coordinates $r_n$ in the trapping potentials. Note that the trapping potential $V(\Pos_n,t)$ only acts in the ground-state, 
as typical magnetic or optical traps designed to trap ground state atoms are ineffective for Rydberg atoms~\cite{anderson:rydbergtrap}.
$\sub{\hat{H}}{c}$ contains the interaction of the atom with the laser (subscript las) and the microwave radiation (subscript rf). Rabi-frequencies are denoted by $\Omega$ and detunings by $\Delta$.
The last term, $\sub{\hat{H}}{int}$, is the interaction between atoms in Rydberg states, where $D_{nl}(\bv{R})=D(|\Pos_n-\Pos_l |)$ describes transition dipole-dipole interactions between an atom in $\ket{s}$ and 
one in $\ket{p}$ and $W^{(ab)}_{nl}=W^{(ab)}(|\Pos_n-\Pos_l |)$ are van-der-Waals (vdW) interactions between atoms $n$ and $l$ with orbital quantum numbers $a,b=s,p$. The vector $\bv{R}=\{r_1,\dots r_{N}\}^T$ contains all atom 
co-ordinates. 

In the following we will use the potentials $D(|\Pos_n-\Pos_l |) = C_3/|\Pos_n-\Pos_l |^3$ and $W(|\Pos_n-\Pos_l |) = C_6/|\Pos_n-\Pos_l |^6$, where $C_3=\pm\mu^2$. The sign of $C_3$ depends on the explicit Rydberg states 
$\ket{s}$ and $\ket{p}$, while $\mu$ is the magnitude of the
transition dipole between the latter. It will be important for the present work that $C_3>0$. The positive sign and the angular independence of $D(|\Pos_n-\Pos_l |)$ can be achieved by choosing the polarization axis of the 
microwave
along the inter-cloud axis \cite{moebius:cradle}, which can reduce the dynamics to just a single selected total angular momentum $m_j$ sublevel of the involved $\ket{p}$ state. Working 
with $s_{1/2}$ and $p_{3/2}$ states ensures $C_3>0$ \cite{park:dipolebroadening}, as required.

\subsection{Numerical solutions}\label{numerics}

For flexible Rydberg systems, where atomic motion and excitation transport due to resonant dipole-dipole interactions affect each other, Tully`s mixed quantum-classical approach~\cite{tully:hopping2,tully:hopping,tully:derivation} is convenient to apply and reliable~\cite{cenap:motion,wuester:cradle,moebius:cradle,moebius:bobbels,moebius:cat}.
Here, we briefly review its core features to explain an extension required for the present work.

We propagate an electronic quantum state $\ket{\Psi(t)}=\sum_{\bv{k}} c_{\bv{k}}(t) \ket{\bv{k}}$ according to the usual time-dependent Schr{\"o}dinger equation (TDSE)
\begin{align}
i \hbar \frac{\partial}{\partial t}\ket{\Psi(t)} &= \sub{\hat{H}}{el}\ket{\Psi(t)},
\label{TDSE}
\end{align}
with electronic Hamiltonian $\sub{\hat{H}}{el} = \sub{\hat{H}}{c} + \sub{\hat{H}}{int}$. The atomic motion is treated through a classical trajectory average with initial conditions drawn from the Wigner function of the lowest harmonic oscillator states within the two wells $A$ and $B$. 

Forces on the atoms arise from the harmonic trap and mutual interactions. Importantly, all these depend on the electronic state. In our algorithm these forces are always calculated as the gradient of \emph{a single} selected Born-Oppenheimer surface $U_p(\bv{R})$. Surfaces are determined from the time-independent Schr{\"o}dinger equation (TISE)
\begin{align}
 \sub{\hat{H}}{el}\ket{\varphi_n (\bv{R})} &= U_n(\bv{R})\ket{\varphi_n (\bv{R})}.
\label{TISE}
\end{align}
In the trajectory average, the surface index $p$ that determines the motion of the atoms is allowed to stochastically switch to another one $l$. The switching probability is determined from the non-adiabatic coupling vector
\begin{align}
\bv{d}_{pl}&=\braket{\varphi_l}{{ \mathbf\nabla}\varphi_p}.
\label{dvec}
\end{align}
Here we will additionally have transitions between different adiabatic surfaces due to the coupling to time-dependent electromagnetic fields. We take these into account with additional stochastic switches, 
due to the non-adiabatic coupling
\begin{align}
t_{pl}&=\braket{\varphi_l}{(\partial/\partial t)\varphi_p}.
\label{tval}
\end{align}
In order to conserve energy in a switch due to \eref{dvec} the velocity is appropriately adjusted along the direction of the non-adiabatic coupling vector $\bv{d}_{pl}$.
~\cite{tully:hopping:veloadjust}. For switches due to \eref{tval} there is no adjustment, as for the duration of coupling pulses energy is not conserved.

\section{Pulsed atom ejection}\label{ejection}

In this section we sequentially describe the required steps for pulsed ejection of EPR entangled atom pairs from the twin atomic clouds sketched in \fref{system}. The simulated sequence here is fast enough to neglect atomic diffusion on the time-scale of atom ejection, hence we set $V(\Pos_n,t)=0$ in \eref{Hamiltonian}.

\subsection{Excitation of  blockade states}\label{exciteblockade}

First we aim to excite a blockade state (also called superatom) in each cloud, so that the total state is $\ket{S}_A\otimes \ket{S}_B$, where $\ket{S}_X=\sum_{n\in X}\ket{\pi_{n(s)}}/\sqrt{N_X}$. We denote by $\ket{\pi_{n(\alpha)}}$ a state where all atoms are in $\ket{g}$, except atom $n$, which is in $\ket{\alpha}$, where $\alpha\in\{s,p\}$. The excitation could be achieved with a Rabi-$\pi$ pulse in $\sub{\Omega}{las}(t)$, as long as the detuning $\Delta$ of the excitation field compensates the van-der-Waals interactions at distance $d$ separating the clouds, thus $\Delta = C_6/d^6$. The detuning then allows us to excite one Rydberg atom in each cloud, despite them being separated by slightly less than the blockade radius, but would still ensure the absence of multiple excitations within one cloud.
Since repeated use of the Rydberg atom source will decrease the number of remaining atoms per cloud $N$, the pulse durations have to be adjusted: The Rabi frequency between $\ket{\bv{g}}$, with all atoms in $\ket{g}$, and $\ket{S}_A$ is $\sub{\Omega}{bl}=\sqrt{N}\sub{\Omega}{las}(t)$. 

Alternatively, repeated application of \emph{identical} excitation sequences can be implemented by using a chirped adiabatic passage, as suggested in \rref{beterov:deterministic}. The frequency (detuning) $\sub{\Delta}{las}(t)$ of the effective laser coupling between $\ket{g}$ and $\ket{s}$ is adjusted from negative detuning to positive detuning in the course of a Gaussian envelope pulse for $\sub{\Omega}{las}(t)$. In this manner the state $\ket{\bv{g}}$ is adiabatically transformed into the state $\ket{S}_X$ in either cloud, regardless of $N$~\cite{footnote:pulseshapes}. The symmetric chirp employed in \rref{beterov:deterministic} would have to be shifted by an offset $\Delta = C_6/d^6$. 

Here, we will employ Gaussian pulses with fixed detuning instead of chirped pulses, resulting in a larger repetition rate. 

\subsection{Excitation of repulsive exciton}\label{exciteexciton}

Next, we transfer the two Rydberg excited atoms to a repulsive exciton state $\ket{\sub{\varphi}{rep}}=(\ket{sp}+\ket{ps})/\sqrt{2}$~\cite{moebius:bobbels} via microwave coupling.
In the same manner that we described in \rref{moebius:cat}, the symmetric repulsive exciton state on a pair of atoms is adiabatically connected with the pair-state $\ket{ss}$, if an initially detuned microwave pulse on the 
$\ket{s}\leftrightarrow\ket{p}$ transitions is chirped from large positive detuning to zero detuning. Since the microwave usually couples with equal phases to all atoms, for inter-atomic distances much less than the wavelength, we 
can only directly access symmetric exciton states in this manner. This explains our choice $C_3>0$ in \sref{hamil}, for which the symmetric state is repulsive.
The r.f.~pulse shapes will be presented later. The final many body state after the r.f.~chirp will be $\ket{\sub{\Psi}{rep}}\equiv(\ket{S}_A\otimes \ket{P}_B + \ket{P}_A\otimes \ket{S}_B)/\sqrt{2}$.

\subsection{Atom ejection}\label{wait}

After one atom in each cloud has been excited to a Rydberg state, and subsequently the Rydberg pair was transferred to a repulsive pair state for resonant dipole interactions, 
these interactions will ultimately push the excited atoms far enough away from their parent clouds to lift the blockade condition and allow the excitation of the next pair of Rydberg atoms. 

\begin{figure}[htb]
\centering
\includegraphics[width=0.99\columnwidth]{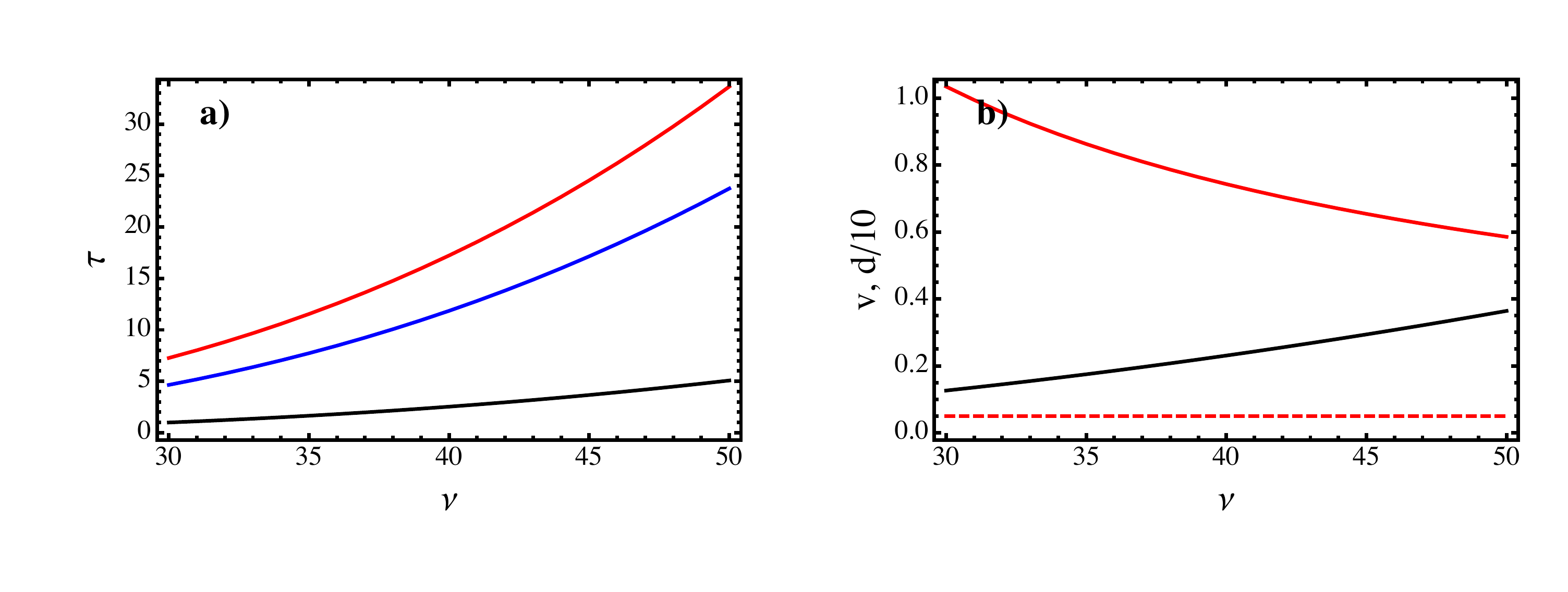}
\caption{(color online) Characteristic features of our scheme as a function of principal Rydberg quantum number $\nu$ for Rb atoms. (a) Acceleration time $\sub{\tau}{acc}$ (black), overall time scale of motion  $\sub{\tau}{acc}+ \sub{\tau}{drift}$ (blue), and life-time $\sub{\tau}{life}$ \cite{beterov:BBR} (red). (b) Final velocity $\sub{v}{fin}$ (red), recoil velocity $\sub{v}{rec}$ (red-dashed) (see \sref{deecitation}) and cloud separation $d$ (black). See footnote \cite{footnote:timescales} for definitions of  the quantities above. For lighter atoms such as Li, the ratio between life time and motional time becomes much larger.
\label{reprate}}
\end{figure}

For a given principal quantum number $\nu$, the timescale of mechanical acceleration can be minimized by placing the two atom clouds as close as possible. We fix the 
inter-cloud distance $d$ to $d=0.65 \sub{r}{bl}$, where $\sub{r}{bl} = (C_6/\sub{\Omega}{las})^{1/6}$ is the van-der-Waals blockade radius. Significantly closer distances would invalidate our effective state model. In order for the last ejected pair of Rydberg atoms to no longer have significant resonant dipole-dipole interactions with subsequently excited atom pairs, they have to travel a distance greater than $d$ away from their parent cloud. 
The time-scale required is given by the distance $d$ divided by the final velocity $\sub{v}{fin}=\sqrt{2 \mu^2/(Md^3)}$, 
where $\mu^2/d^3$ is the initial dipole-dipole interaction energy. Inserting the relation between $d$ and the principal quantum number $\nu$ which arises from $d \sim \sub{r}{bl}$ we find that the motional time
scales like $\nu^{31/12}$~\cite{footnote:timescales}. Fast repetition rates thus favor \emph{smaller} principal quantum numbers. In \fref{reprate} we show the times required for acceleration and motion, 
compared to the pair life-time, as a function of the principal quantum number $\nu$.
%

\subsection{Atom de-excitation}\label{deecitation}

The measurement of the entanglement of EPR pairs, as discussed in \sref{entanglement}, may benefit from the atoms being in Rydberg state by using state 
selective field ionization~\cite{walz:stateselective,reinhard:stateselective,preclikova:stateselective} to infer pseudo-spin states. 
Alternatively, if one prefers long lived entanglement, one can de-excite the Rydberg atoms to two different long lived ground states $\ket{s}\rightarrow\ket{g}$, $\ket{p}\rightarrow\ket{h}$ (see also \cite{wuester:dressing}). These could be members  of the ground hyperfine multiplet $\ket{F,m_F}$, where $F$ is the total angular momentum and $m_F$ the associated magnetic quantum number, for example $\ket{g}=\ket{1,-1}$ and $\ket{h}=\ket{2,1}$. A further advantage of controlled de-excitation is to avoid uncontrolled atomic recoil due to spontaneous emission from the atoms ejected by the source, for cases where a directed atomic beam is desirable. If the atom incurs a recoil, it gains a velocity $\sub{v}{rec}=E(\nu)/(Mc)$ of about $0.04$m/s for Rydberg states of Rb. In the expression for $\sub{v}{rec}$, the speed of light is $c$ and the energy difference to the ground-state $E(\nu)\approx13.6$eV. Compared to a directed drift velocity due to dipole-dipole interactions, of the order of $0.5$m/s the recoil is thus relatively small.

\subsection{Minimal model demonstration}\label{simulations}

The least number of atoms with which one can demonstrate essential features of the atom pair production is $N_A=N_B=2$. This is done in the following, modelling Rydberg excitation and acceleration for $2$ Rubidium atoms per cloud, for $d=3\mu$m, $\sigma=0.3\mu$m, $\nu=42$, which results in $\mu=1715$ atomic units. 
Atoms are subjected to a laser pulse followed by an r.f.~chirp, separated by $1\mu$s wait time. The total electronic wave function in this section will be written as
\begin{align}
\ket{\Psi(t)}&=\sum_{n_1,n_2:n_3,n_4}c_{n_1,n_2:n_3,n_4}(t) \ket{n_1 n_2:n_3 n_4},
\label{fourbodyelectronic}
\end{align}
with $n_i \in \{g,s,p\}$. The electronic basis states $\ket{n_1 n_2:n_3 n_4}$ are a more explicit notation for the $\ket{\bv{k}}$ in \sref{hamil}.
Indices to the left of the colon label atom one and two, in cloud A, those to the right atom three and four, in cloud B. We take cut-off vdW interactions into account to ensure a blockade for the propagation of the electronic states in Tully's algorithm, however, for the mechanical motion of atoms vdW forces can be neglected here~\cite{footnote:vdW}.

\begin{figure}[htb]
\centering
\includegraphics[width=0.99\columnwidth]{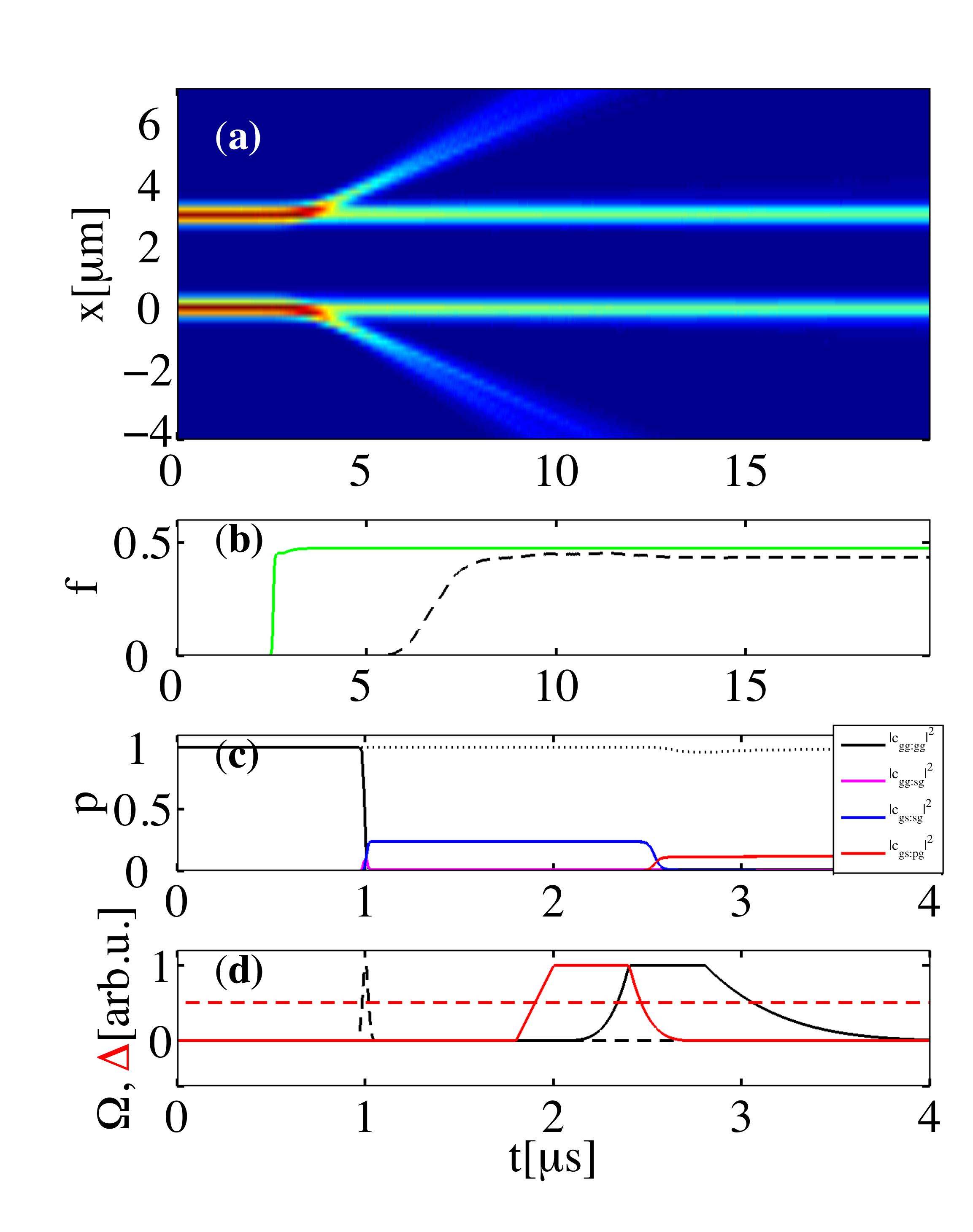}
\caption{(color online) Ejection of an entangled Rb atom pair from two harmonic traps, using a quantum-classical model with 1024 stochastic trajectories. The principal quantum number is $\nu=55$. (a) Total atomic density. 
(b) Fraction $f$ of atoms ejected from the clouds (black), and expected ejection fraction based on the population in the repulsive exciton state $|\braket{\sub{\Psi}{rep}}{\Psi(t)}|^2$ (green).
(c) Population in the various electronic states during the first laser and r.f. pulse. For definitions of $c_{n_1,n_2:n_3,n_4}$ see \eref{fourbodyelectronic}. 
States where atoms within one cloud are permuted or the clouds swapped have identical populations to those listed in the legend, by symmetry.
The dotted line is the total population in all states shown in the legend, and their permutations. 
(d) Laser (dashed) and microwave (solid) pulse shapes (normalised arbitrarily, for more details see \cite{footnote:pulseshapes}). The detuning is shown in red and the Rabi-frequencies in black.
\label{simulation}}
\end{figure}

The results are shown in \fref{simulation}. The total atomic density from the trajectory average shows the two parent clouds and the ejection of an atom pair.
While precisely one atom is ejected from each cloud, the underlying quantum state is 
 a superposition of all combinations with either of the two atoms from a cloud ejected \cite{moebius:bobbels}. 
The population in electronic states, shown in panel (c), indicates almost perfect conversion, first from the state $\ket{\bv{g}}$, with all atoms in $\ket{g}$, to the blockade state via the chirped laser coupling, then from a blockade state in each cloud to a many-body repulsive exciton. This state is a superposition where each possible pair of one atom from cloud $A$ and one from cloud $B$ is with equal probability in the repulsive exciton state $\ket{\sub{\varphi}{rep}}=(\ket{sp} + \ket{ps})/\sqrt{2}$. 

The transitions in the electronic space lead to the correct fraction of ejected atoms, as shown in panel (b), hence adding non-adiabatic transitions according to \eref{tval} to the surface hopping algorithm is appropriate.

Finally the figure also shows the shape of the laser and r.f.~pulses and their respective frequencies, precise parameters are listed in \cite{footnote:pulseshapes}.

As explained in \cite{moebius:bobbels}, the physics presented in this section remains unchanged if we begin with many more atoms than the $N=4$ modelled here, as long as the assumption of a full blockade of each cloud, but completely lifted blockade from one cloud to the other is fulfilled. Repeated application of identical outcoupling sequences as in \fref{simulation} then leads to a pulsed beam of single atoms, pairwise EPR entangled between two different beams. 

Let us complete this section with an outlook on possible variants of atom ejection using dipole-dipole interactions. {\it Variant (i):} Instead of exciting $\ket{s}$ Rydberg states in both clouds and subsequently accessing $\ket{p}$ via microwave transitions, one could also excite atoms in cloud $A$ to $\ket{s}$ and in cloud $B$ to $\ket{p}$. The resulting exciton state is a superposition of repulsive and attractive dynamics, so that the atoms are ejected \emph{towards each other} in 50\% of the cases. 
Post selecting only the repulsive ones yields the same entanglement structure as in our original scheme. {\it Variant (ii):} One could excite $\ket{s}$ Rydberg states in both clouds, and subsequently admix population from neighboring $\nu$ manifolds 
via a $\ket{ss}\leftrightarrow\ket{pp'}$ F{\"o}rster resonance~\cite{altiere:foerster,vogt:foerster,ryabtsev:foerster}. 
Accessing the exciton proceeds via ramps of a static electric field here, instead of microwave pulses. Atoms in this case form entangled two-body states of the schematic form ${\cal N}(\ket{ss} + c_p \ket{pp'})$ where ${\cal N}$ is a normalization factor and $|c_p|^2<1$ depends on the F{\"o}rster defect. 

%
\section{Entanglement measurements}
\label{entanglement}

Concentrating on the ejected entangled atoms, the setup shown in \fref{system} largely resembles the 
variant of the EPR paradox proposed by Bohm and Aharonov~\cite{aharonov:bohm:EPR}. 
The Bohm and Aharonov scheme is based on the decay of a spin $0$ particle into a pair of spin $1/2$ particles in a spin singlet state 
{of the form $\ket{\Psi_S}=(\ket{\uparrow\downarrow}-\ket{\downarrow\uparrow})/\sqrt{2}$}.  Let us denote these particles by $a$ and $b$.
{The measurable correlation between spin projections along axis $\bv{a}$ for particle $a$ and $\bv{b}$ for particle $b$ is expressed as
\begin{align}
C_S^{(\bv{a},\bv{b})}&=\langle \Psi_S\,|\, \boldsymbol{\sigma}_a \cdot \bv{a}   \:\:\:  \boldsymbol{\sigma}_b \cdot \bv{b}\,|\,\Psi_S\rangle = -  \bv{a}\cdot \bv{b}.
\label{EPRcorrel}
\end{align}}
Here, ${\boldsymbol \sigma}_j$ is a vector of Pauli spin matrices acting in the spin space of atom $j$ and $\bv{a}\cdot \bv{b}$ denotes the scalar product between the vectors $\bv{a}$ and $\bv{b}$.

\begin{figure}[htb]
\centering
\includegraphics[width=0.99\columnwidth]{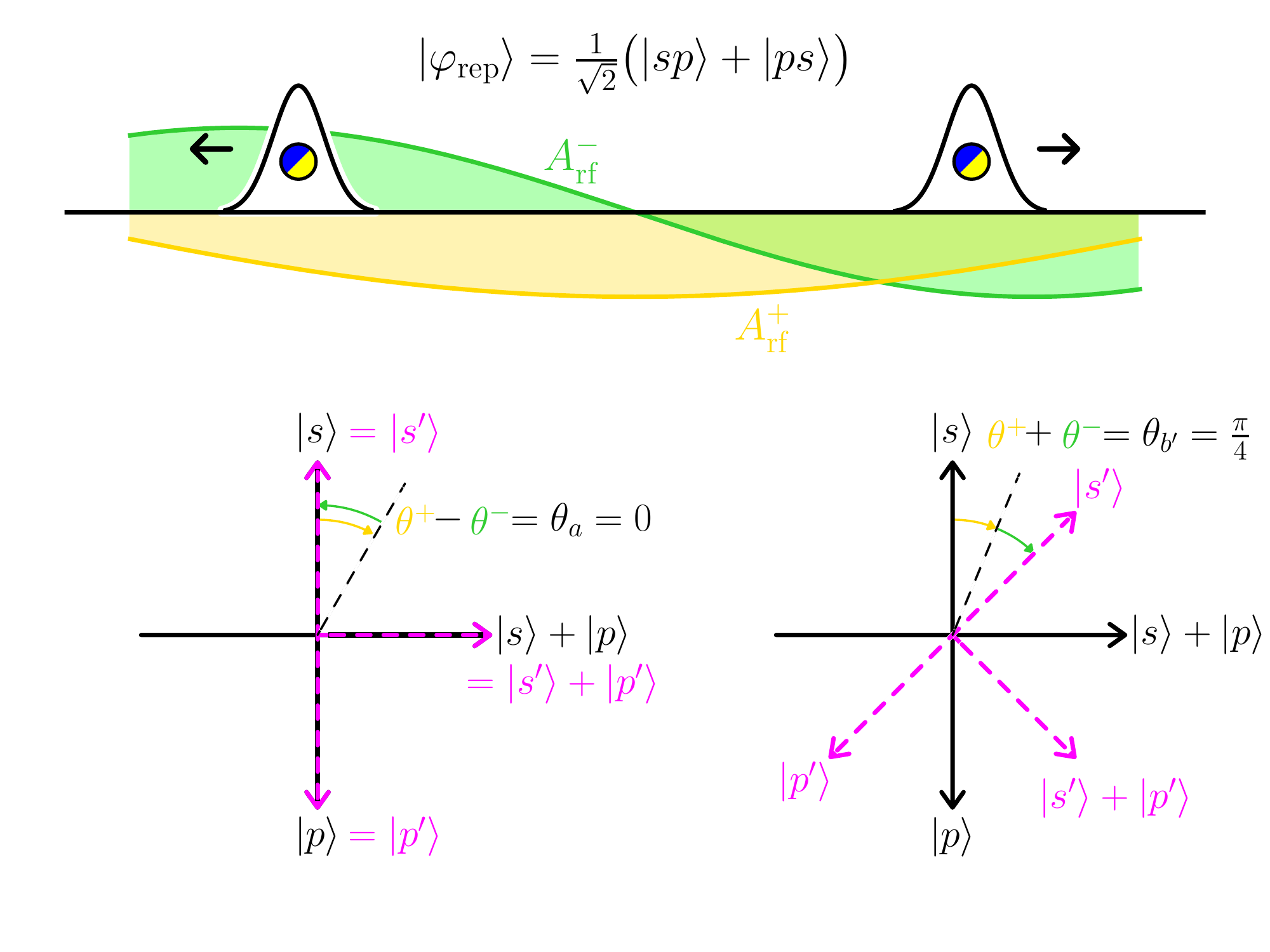}
\caption{(color online) Measurement scheme for atomic pseudospin EPR correlations. (top) Atoms are ejected towards different sides of a microwave node, realizing either same or opposite sign Rabi coupling between $\ket{s}$ $\ket{p}$. 
(bottom) Cuts through the Bloch sphere. Combining symmetric and antisymmetric coupling, we can rotate the measurement basis differently for atoms $a$ and $b$. 
\label{measurement_scheme}}
\end{figure}
By identifying electronic states with spin states according to $\ket{s}\rightarrow\ket{\uparrow}$, $\ket{p}\rightarrow\ket{\downarrow}$ (or if we work with de-excited ground states, \sref{deecitation}, $\ket{g}\rightarrow\ket{\uparrow}$, $\ket{h}\rightarrow\ket{\downarrow}$), we can view the ejected atoms as a coupled spin-$1/2$ system. However, straightforward r.f.~excitation of the repulsive exciton requires it to have the 
form $\ket{\sub{\varphi}{rep}}=(\ket{sp}+\ket{ps})/\sqrt{2}$, 
which in the spin picture corresponds to a member of the triplett. In this triplet state $\ket{\Psi_T}=(\ket{\uparrow\downarrow}+\ket{\downarrow\uparrow})/\sqrt{2}$ we obtain \cite{barut:EPRtriplet}
\begin{align}
C_T^{(\bv{a},\bv{b})}&=\bra{\Psi_T} \boldsymbol{\sigma}_a \cdot \bv{a}   \:\:\:  \boldsymbol{\sigma}_b \cdot \bv{b}\ket{\Psi_T} =  \bv{a}\cdot \bv{b} - 2a_z b_z
\label{EPRcorrel_triplett}
\end{align}
{for the correlation}, where $a_i$, $b_i$ are the cartesian components of the vectors $\bv{a}$ and $\bv{b}$.

In both, singlet and triplet cases, essential nonclassical features of entanglement are evident if one can violate a Bell inequality \cite{bell:theorem}, 
for example the CHSH form~\cite{chsh:inequality}
\begin{align}
|C_T^{(\bv{a},\bv{b})} + C_T^{(\bv{a},\bv{b'})} + C_T^{(\bv{a'},\bv{b})} - C_T^{(\bv{a'},\bv{b'})}|&\leq2,
\label{CHSHineq}
\end{align}
for any choice of axes $\bv{a}$, $\bv{a'}$, $\bv{b}$, $\bv{b'}$.
All classical, realistic, local, hidden variable theories would have to fulfill \eref{CHSHineq}. 

{
If we wish to realize a violation of Eq.~(\ref{CHSHineq}) using the emitted  pair of Rydberg atoms, we thus should be able to independently control
the measurement axes $\bv{a}$, $\bv{a'}$, $\bv{b}$, $\bv{b'}$. However, as we can only measure the angular quantum number of the outgoing Rydberg atoms - corresponding in the spin-picture to a measurement of the $z$-component - we cannot directly access the non-classical region. 
This problem can be solved by coupling the $\ket{s}$ and $\ket{p}$ states for each of the emitted atoms to each other prior to the Rydberg state measurement.
Let $V_c$ ($c=a,b$) denote the respective coupling strengths and $\tau$ the coupling duration. The corresponding Hamiltonian for a single atom in the basis ($\ket{s}$, $\ket{p}$) thus reads
}
{
\begin{equation}\label{Hcouple}
H_c=\left(\begin{array}{cc}
0 & V_c \\
V_c & 0
\end{array}\right),
\end{equation}
}
{
and the time evolution operator is given by
\begin{equation}\label{Hcouple}
U_c=\exp\left(-iH_c\tau\right)=\left(\begin{array}{cc}
\cos(\theta_c/2) & -i\sin(\theta_c/2) \\
-i\sin(\theta_c/2) & \cos(\theta_c/2)   
\end{array}\right),
\end{equation}
where $\theta_c=2V_c\tau$. Note that the coupling is applied to each of the two atoms individually, hence we can write 
the full time evolution operator for the two-atom system as $U=U_a\otimes U_b$. Consider now some arbitrary initial two-particle state $\ket{\chi}$.
One can easily verify that a $z$-component measurement for this state {\it after} the coupling period (at $t=\tau$) is equivalent to
a correlation measurement such as in Eqs.~(\ref{EPRcorrel}) and~(\ref{EPRcorrel_triplett}) {\it before} the coupling period (at $t=0$),
where the two axes $\bv{c}=\bv{a},\bv{b}$ are given by 
\begin{align}
\bv{c} = [0,\sin(\theta_c),\cos (\theta_c)]^T,
\label{measureaxis}
\end{align}
i.e., the following equality holds:
\begin{align}
\langle \chi\,|\, \boldsymbol{\sigma}_a \cdot \bv{a}   \:\:\:  \boldsymbol{\sigma}_b \cdot \bv{b}\,|\,\chi\rangle = \langle \chi\,|\, U_a^{\dagger}\,\sigma_a^z \:\:\:  \sigma_b^z\,U_b \,|\,\chi\rangle.
\end{align}
Hence,} 
the coupling effectively allows a rotation of the measurement axes.  For axes given in \eref{measureaxis}, we can rewrite \eref{CHSHineq} as
\begin{align}
&|\cos{(\theta_{\bv{a}} + \theta_{\bv{b}})} + \cos{(\theta_{\bv{a}} + \theta_{\bv{b'}})} 
\CR
&+ \cos{(\theta_{\bv{a'}} + \theta_{\bv{b}})} - \cos{(\theta_{\bv{a'}} + \theta_{\bv{b'}})}|\leq2.
\label{CHSHineq2}
\end{align}
Maximal violation of \bref{CHSHineq2} is achieved for example by $\theta_{\bv{a}}=0$, $\theta_{\bv{a'}}=\pi/2$, $\theta_{\bv{b}}=-\pi/4$ and $\theta_{\bv{b'}}=\pi/4$.

The required coupling between $\ket{s}$ and $\ket{p}$ can be provided with microwaves. These would, however, usually couple symmetrically to atoms $a$ and $b$, 
which would not allow us to violate \bref{CHSHineq}. To obtain independent control of the pseudo-spin measurement axes of the two ejected atoms, we require an 
antisymmetric microwave coupling, that could arise on different sides of a field node in an r.f.~resonator
as sketched in \fref{measurement_scheme}. Combinations of symmetric ($A_{rf}^{(+)}$) and antisymmetric ($A_{rf}^{(-)}$) 
pulses then can realize the axes necessary for violating the Bell inequality. The example $\theta_{\bv{a}}=0$ with 
$\theta_{\bv{b'}}=\pi/4$ is sketched in \fref{measurement_scheme}. Other ways to achieve independent measurement axes on atoms $a$ and $b$, 
is to shelve one of the atoms into ground states $\ket{g}$ and $\ket{h}$ for the duration of microwave coupling, and thus coupling $\ket{s}$ to $\ket{p}$ for the remaining Rydberg atom only, or employing a double off-resonant Raman transition that can address an individual atom through laser focusing.

\section{Conclusions and outlook}

A pair of ultra-cold atom clouds, each confined tighter than the Rydberg blockade radius and separated by a distance of the order of the blockade radius, can emit pairs of EPR correlated Rydberg or ground-state atoms on demand.
The ejected atoms can be shown to violate a Bell inequality in the Rydberg state space with standard methods. Additionally, the setup provides a pulsed single-atom source. The quantum-classical hybrid method used in this article allows for an elegant way to model blockade, Rydberg excitation and acceleration by state dependent dipole-dipole forces in a single framework.

\acknowledgments
We acknowledge useful comments from Thomas Pohl and EU financial support received from the Marie Curie Initial Training Network (ITN) "COHERENCE".


\begin{thebibliography}{57}
\bibitem[{foo({\natexlab{a}})}]{footnote:pulseshapes}
\bibinfo{note}{The laser coupling has temporal profiles
  $\sub{\Omega}{las}(t)=\sum_n \sub{A}{las}
  \exp[-(t-\sub{t}{0,las,n})^2/(2\sub{\tau}{las}^2)]$ and
  $\sub{\Delta}{las}(t)=\sum_n \theta(\sub{\tau}{det} - |t-\sub{t}{0,las,n}|)
  [\sub{\Delta}{las,-} +
  (\sub{\Delta}{las,+}-\sub{\Delta}{las,-})(t-\sub{t}{0,las,n}-\sub{\tau}{det})/(2\sub{\tau}{det})]$,
  where $\sub{\Delta}{las,$\pm$}$ are the maximal and minimal detuning.
  Parameters for the laser coupling were $\sub{A}{las}=(2\pi)2$MHz,
  $\sub{t}{0,las,0}=9\mu$s, $\sub{t}{0,las,0}=59\mu$s,
  $\sub{\tau}{las}=2.5\mu$s, $\sub{\tau}{det}=\sub{\tau}{las}$,
  $\sub{\Delta}{las,-}=-(2\pi)8.4$MHz and $\sub{\Delta}{las,+}=(2\pi)3.3$MHz.
  R.f.~pulse shapes in the next section are more complicated. They are
  displayed in \frefp{simulation}{d,e}, where the maximal detuning (Rabi
  frequency) was $(2\pi)500$ MHz ($(2\pi)20$ MHz).}
\bibitem[{foo({\natexlab{b}})}]{footnote:timescales}
\bibinfo{note}{We assume an acceleration time
  $\sub{\tau}{acc}=\sqrt{2M/3}d^{5/2}/\mu$, and a drift time
  $\sub{\tau}{drift}=L/\sub{v}{fin}$, $\sub{v}{fin}=\sqrt{2\mu^2/(Md^3)}$, all
  from Newton's equations. The drift length of the atoms $L$ is dependent on
  the experimental setup and set to $L\approx 2d$ here. We estimate the drift
  velocity as $ \sub{v}{fin}= a(d) \sub{\tau}{acc}$, where $a(d)=3\mu^2/d^4/M$
  is the initial acceleration. In our estimates we use the scaling law
  $\mu=\mu_0 \nu^2$ for the transition dipole, with $\mu_0=0.94/\sqrt{3}$
  a.u.~,extrapolating from the reference value $\mu=966/\sqrt{3}$ a.u.~for
  $\nu=32$ \cite{westermann:transfer}. Lifetimes of a Rydberg pair state
  $\ket{sp}$, including the effect of black-body radiation, are calculated as
  $\sub{\tau}{life} = (1/\tau_s + 1/\tau_p)^{-1}$, with
  $\tau_{s/p}=\tau_{0s/p}\nu^3$ and $\tau_{0s}=0.48$ns, $\tau_{0p}=0.61$ns
  \cite{beterov:BBR}.}
\bibitem[{foo({\natexlab{c}})}]{footnote:vdW}
\bibinfo{note}{At the given distance, the dipole-dipole force for the pair
  state $\ket{sp}$ is about 20 times stronger than the van-der-Waals force for
  the pair state $\ket{ss}$, which is created in the first step of our scheme
  (cf. Section III A). In the time span between the creation of the $\ket{ss}$
  state and the excitation of the repulsive exciton, the van-der-Waals-induced
  motion can shift the atoms by about $\sim 0.03\,\mu{\rm m}\approx 0.1\sigma$,
  which is insignificant and hence neglected in our simulations.}
\end{thebibliography}
\end{document}